\documentclass[aps]{revtex4}
\usepackage{graphicx,psfrag}
\def\bc{\begin{center}}
\def\ec{\end{center}}

\def\beq{\begin{equation}}
\def\eeq{\end{equation}}

\def\br{{\bf r}}

\begin{document}

\title{Short note on the density of states in 3D Weyl semimetals
}

\author{K. Ziegler and A. Sinner}
\affiliation{
Institut f\"ur Physik, Universit\"at Augsburg,
D-86135 Augsburg, Germany}
\date{\today}

\begin{abstract}
The average density of states in a disordered three-dimensional Weyl system is discussed in the
case of a continuous distribution of random scattering. Our results clearly indicate that
the average density of states does not vanish, reflecting the absence of a critical point
for a metal-insulator transition. This calculation supports recent suggestions of an avoided
quantum critical point in the disordered three-dimensional Weyl semimetal. However, the
effective density of states can be very small such that the saddle-approximation with
a vanishing density of states might be valid for practical cases.
\end{abstract}

\maketitle

\section{Introduction}

The existence of a metal-insulator transition in disordered three-dimensional (3D) Weyl semimetals 
has been debated in the recent literature 
\cite{huse13,brouwer14,gurarie14,pixley16,pixley16a,pixley16b,ziegler16,pixley17,sbierski17,sinner17,1805}. 
It is closely
related to the question, whether or not the average density of states (DOS) at the spectral 
node vanishes below some critical disorder strength. The self-consistent Born approximation 
provides such a critical value with a vanishing DOS for weak disorder. It has been argued 
that rare regions of the random distribution may lead to a non-vanishing average DOS, though \cite{huse13}. 
This was supported by recent
numerical studies based on the T-matrix approach, which gives an exponentially small
DOS \cite{pixley17} but was questioned in a recent study based on an instanton solution \cite{1805}.
In this short note we show that, depending on the type and strength,
a continuous distribution of disorder can create a substantial average DOS at the 
spectral node in 3D Weyl systems. This requires at least two impurities to create a resonant
state between these impurities. A single impurity or a single instanton does not
contribute to the spectral weight at the Weyl node, though, in accordance with the arguments in 
Ref. \cite{1805}. 
This supports the picture of an avoided quantum critical point in the presence of a 
distribution of impurities, as advocated in Ref. \cite{pixley17}.

\section{Model}

The 3D Weyl Hamiltonian for electrons with momentum ${\vec p}$ is expanded 
in terms of Pauli matrices $\tau_j$ ($j=0,1,2,3$; $\tau_0$ is the $2\times2$ unit matrix) 
as $H=H_0-U\tau_0$, where 
\beq
H_0=v_F{\vec\tau}\cdot{\vec p}
\ \ \ {\rm with}\ \ {\vec\tau}=(\tau_1,\tau_2,\tau_3)
\ .
\label{ham00}
\eeq
$v_F$ is the Fermi velocity and
$U$ is a disorder term, represented by a random potential with mean $\langle U\rangle=E_F$ (Fermi energy)
and variance $g$. 
The average Hamiltonian $\langle H\rangle=H_0-E_F\tau_0$ 
generates a spherical Fermi surface with radius $|E_F|$, and
with electrons (holes) for $E_F>0$ ($E_F<0$). Physical quantities are expressed in such units that $v_F\hbar=1$.

The DC limit $\omega\to0$ of the conductivity of 3D Weyl fermions depends only on the scattering rate $\eta$ 
and the Fermi energy $E_F$
\cite{ziegler16}:
\beq
\sigma(\eta,E_F)=2\frac{e^2}{h}\eta^2\int_0^\lambda\frac{(\eta^2+k^2)^2+E_F^2(2\eta^2+2k^2/3+E_F^2)}
{[(\eta^2-E_F^2+k^2)^2+4\eta^2E_F^2]^2} \frac{k^2dk}{2\pi^2}
\label{cond2}
\eeq
with momentum cut-off $\lambda$. 
At the node ($E_F=0$) the DC conductivity in Eq. (\ref{cond2}) is reduced to the expression
\beq
\sigma =2\frac{e^2}{h}\eta^2\int_0^\lambda \frac{k^2}{(\eta^2+k^2)^2}\frac{dk}{2\pi^2}
=\frac{e^2\eta}{2\pi^2h}\left[\arctan(1/\zeta)-\frac{\zeta}{1+\zeta^2}\right] \ \ \ 
(\zeta=\eta/\lambda)
\ ,
\label{cond000}
\eeq
which becomes for $\lambda\gg\eta$
\beq
\sigma\sim \frac{ e^2}{4\pi h}\eta
\ .
\label{cond001}
\eeq
The last result was also derived by Fradkin some time ago \cite{fradkin86a}. 
In contrast to the 2D case, where $\sigma=e^2/\pi h$, the 3D case gives a linearly 
increasing behavior with respect to the scattering rate. 

The results in (\ref{cond2}) -- (\ref{cond001}) clearly indicate that a metal-insulator transition 
in disordered 3D Weyl systems is directly linked to the scattering rate $\eta$. The latter describes 
the broadening of the poles of the one-particle Green's function and is proportional to 
the average DOS
\beq
\rho_{\br}(E_F)=\lim_{\epsilon\to 0}\frac{1}{\pi}Im\left[{\bar G}_{\br\br}(-i\epsilon))\right]
\ , \ \ \ 
{\bar G}(-i\epsilon)
=\langle(H_0-U\tau_0-i\epsilon)^{-1}\rangle
\ ,
\label{av_dos}
\eeq
where is ${\bar G}_{\br\br}$ is the diagonal element of ${\bar G}$
with respect to space coordinates.
The self-consistent Born approximation \cite{fradkin86a,ziegler16} at the node $E_F=0$ reads 
\beq
\eta=\eta I
\\ \ \ {\rm with}\ \  
I=\gamma\left[
\lambda-\eta\arctan(\lambda/\eta)\right] 
\eeq
for the effective disorder strength $\gamma=g/2\pi^2$.
There are two solutions, namely $\eta=0$ and a solution with $\eta\ne0$,
which exists only for
sufficiently large $\gamma$. Moreover, $\eta$ vanishes continuously as we reduce $\gamma$. 
For $\eta\sim 0$ we obtain the linear behavior
\beq
\eta\sim\frac{2\lambda}{\pi}(\gamma\lambda -1)
\ ,
\label{scba2}
\eeq
where $\gamma_c=1/\lambda$ appears as a critical point with $\eta=0$ for $\gamma\le \gamma_c$
and $\eta>0$ for $\gamma>\gamma_c$.

\begin{figure*}[t]
\includegraphics[width=8cm]{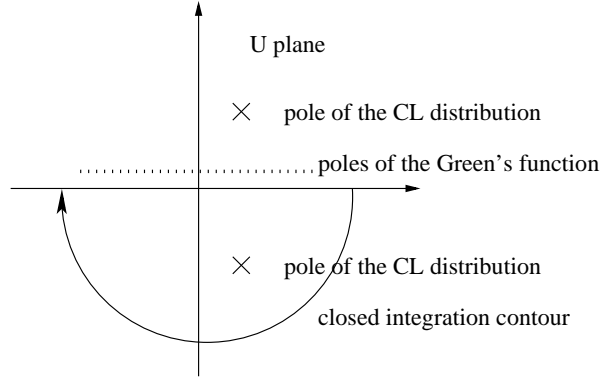}
\caption{
Poles of the one-particle Green's function and the Cauchy-Lorentz distribution. 
The contour of the $U_\br$--integration encloses only one pole of the Cauchy-Lorentz 
distribution but not the other poles.  
}
\label{fig:c-plane}
\end{figure*}

\section{Average density of states}

\subsection{Few impurities: Lippmann-Schwinger equation}

At the node $E_F=0$ the pure DOS $\rho_{0;\br}(E_F=0)$ vanishes. However,
a few impurities have already a significant effect on the local DOS: 
Assuming an impurity potential $U_N$ on $N$ sites, we use the identity (lattice version of the 
Lippmann-Schwinger equation) 
\beq
(G_0^{-1}-U_N)^{-1}=G_0 + G_0({\bf 1} - U_N P_NG_0P_N)^{-1}_NU_N G_0
\ ,
\label{LSE}
\eeq
where $P_N$ is the projector on the impurity sites and $(...)^{-1}_N$ is the inverse 
on the impurity sites. Although $\rho_{0;\br}(E_F=0)$ vanishes,
the second term on the right-hand side of Eq. (\ref{LSE}) can contribute with the poles of 
$({\bf 1} - U_N P_NG_0P_N)^{-1}_N$
to the DOS. These poles are ``rare events'' and require a fine-tuning of the impurity potential, 
whereas the generic case of a general $U_N$ would still have a vanishing DOS. In a realistic situation
the number of impurities is macroscopic with a non-zero density in the infinite system. Then
the identity (\ref{LSE}) cannot be used for practical calculations and we have to average over 
many impurity realizations.
This leads to the average Green's function of Eq. (\ref{av_dos}), which will be calculated subsequently.

\subsection{One vs. two impurities}

The Green's function $G_0$ of the system without impurities reads 
\beq
G_{0,\br}(-i\epsilon)=\frac{1}{|{\cal B}|}\int_{\cal B}\frac{e^{i{\vec k}\cdot\br}}{\epsilon^2+k^2}
\left(i\epsilon\tau_0+{\vec k}\cdot {\vec \tau}\right)d^3k
\equiv i\epsilon\gamma_0\tau_0+{\vec \gamma}\cdot{\vec \tau}
\ ,
\label{1pgf}
\eeq
where ${\cal B}$ is the Brillouin zone of the underlying lattice and 
\[
\gamma_0=\frac{1}{|{\cal B}|}\int_{\cal B}\frac{e^{i{\vec k}\cdot\br}}{\epsilon^2+k^2}d^3k
\ ,\ \ \ 
\gamma_j=\frac{1}{|{\cal B}|}\int_{\cal B}\frac{e^{i{\vec k}\cdot\br}k_j}{\epsilon^2+k^2}d^3k
\ \ \ (j=1,2,3)
\ .
\]
Then the diagonal element $G_{0,0} =i\epsilon \gamma\tau_0$ vanishes with $\epsilon\sim0$. 
This implies that for a single impurity there is no bound 
state at finite impurity strength $U_\br$, since in the impurity term of the Lippmann-Schwinger
equation (\ref{LSE}) the $2\times2$ matrix
\beq
({\bf 1} - U_\br P_\br G_0P_\br)^{-1}=\frac{1}{1-i\epsilon \gamma_0 U_\br}\tau_0
\label{2x2}
\eeq
has a pole at $U_\br\sim \infty$.
The latter reflects the statement that a potential well in 3D Weyl semimetals does never generate 
spectral density at zero energy \cite{1805}.
For two impurities, though, there is a resonant inter-site bound state between the impurities, 
since $G_{0,\br-\br'}$ ($\br'\ne \br$) does not vanish for $\epsilon\to 0$ but decays with a 
power law for $|\br-\br'|$ due to the Pauli matrix coefficients $\gamma_j$ in Eq. (\ref{1pgf}):
\beq
({\bf 1} - U P_{\{\br,\br'\}}G_0P_{\{\br,\br'\}})^{-1}
=\pmatrix{
1-i\epsilon\gamma_0U_\br & 0 & -U_\br\gamma_3 & -U_\br(\gamma_1-i\gamma_2) \cr
0 & 1-i\epsilon\gamma_0U_\br & -U_\br(\gamma_1+i\gamma_2) & U_\br\gamma_3 \cr
-U_{\br'}\gamma_3 & -U_{\br'}(\gamma_1-i\gamma_2) & 1-i\epsilon\gamma_0U_{\br'} & 0 \cr
-U_{\br'}(\gamma_1+i\gamma_2) & U_{\br'}\gamma_3 & 0 & 1-i\epsilon\gamma_0U_{\br'} \cr
}^{-1}
\ .
\eeq
The degenerate eigenvalues of this matrix 
\beq
\frac{1}{1-i\epsilon\gamma_0(U_{\br}+U_{\br'})/2\pm \sqrt{U_{\br}U_{\br'}
(\gamma_1^2+\gamma_2^2+\gamma_3^2)-\epsilon^2\gamma_0^2(U_{\br}-U_{\br'})^2/4}}
\eeq
have poles for finite $U_{\br}$, $U_{\br'}$.
Thus, the corresponding bound states contribute with a non-vanishing density of states. 
In the remainder of the paper this result will be generalized to multiple impurities with corresponding 
resonant bound states.

\subsection{Distribution with simple poles}
\label{sect:CLD}

From here on we consider a continuous distribution of the disorder potential $U$ with
$\prod_\br P(U_\br)dU_\br$ and average one-particle Green's function
\beq
{\bar G}(-i\epsilon)=\int (H_0-U\tau_0-i\epsilon)^{-1}\prod_\br P(U_\br)d U_\br
\ .
\eeq
For $\epsilon>0$ the one-particle Green's function $(H_0-U\tau_0-i\epsilon)^{-1}$ has poles 
for $U_\br$ on the upper complex half-plane.
Assuming that the distribution density $P(U_\br)$ has isolated poles in the lower complex half-plane,
the Cauchy integration can be applied by closing the integration along the real axis in the
lower complex half-plane,
as depicted in Fig. \ref{fig:c-plane}. The simplest realization is the Cauchy-Lorentz distribution
\beq
P_{CL}(U_\br)=\frac{1}{\pi}\frac{\eta}{(U_\br-E_F)^2+\eta^2}
\ ,
\eeq
which gives
\beq
{\bar G}(-i\epsilon)
=(H_0-(E_F+i\epsilon+i\eta)\tau_0)^{-1}
\ .
\eeq
The average DOS then reads
\beq
\rho_{\br}(E_F)=\frac{\eta}{\pi}
[(H_0-E_F\tau_0)^2+\eta^2\tau_0]^{-1}_{\br\br}
\ .
\eeq
The Cauchy-Lorentz distribution has an infinite second moment (i.e., $g$ is infinite). 
A distributions with a finite second moment can be created from the differential of the Cauchy-Lorentz
distribution with respect to $\eta$. 
Many distributions, like the popular Gaussian distribution
\beq
P_G(U_\br)=\frac{1}{\sqrt{\pi g}}e^{-(U_\br-E_F)^2/g}
\ ,
\eeq
do not have a simple pole structure, though. Then another approach can be applied to show that 
there is a non-vanishing average DOS.

\begin{figure*}[t]
\includegraphics[width=4cm]{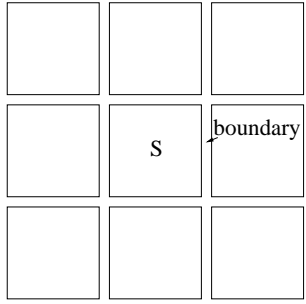}
\caption{
Dividing the system into cubes $\{ S\}$ of size $|S|$ with boundary $\partial S$.
}
\label{fig:cubes}
\end{figure*}

\subsection{Distribution without simple poles}
\label{sect:lower_bound}

Now we only assume that the distribution of $U_\br$ is continuous. Then the path
of integration can also be deformed away from the poles of the Green's function
to obtain a similar result as in the case of simple poles. The calculation would be 
more complex, though. Therefore, we use a different approach, whose main idea is to
divide the system into cubes $\{ S\}$ of finite identical size (cf. Fig. \ref{fig:cubes}).
Then we estimate (i) the average DOS inside an isolated cube and (ii) the 
contribution of the boundary $\partial S$ between the cubes. This approach was used for a periodic
lattice \cite{ledermann44}, for a random tight-binding model with symmetric Hamiltonian \cite{wegner81}
and for two-dimensional Dirac fermions with random mass \cite{ziegler87}. Later it was applied to 
S-wave superconductor with random order parameter \cite{ziegler88}, and to D-wave superconductor with 
random chemical potential \cite{ziegler98}.

For the average local DOS
\beq
{\bar \rho}_\br=\int\rho_\br(U) \prod_{\br}P(U_\br)dU_\br
\eeq
we obtain from the estimation procedure with steps (i) and (ii) the inequality 
(cf. Supplemented Materials)
\beq
\sum_{\br\in S}{\bar\rho}_\br\ge
\inf_{\{-a\le U_\br'\le a\}}\left[\int_{-v}^v
\sum_{\br \in S}\rho_{S,\br}(U'+E)dE \inf_{-v\le w\le v}\prod_{\br\in S}P(U_\br'+w)\right]
-{\bar P}_S|\partial S|
\ ,
\label{bound}
\eeq
where $|S|$ ($|\partial S|$) is the number of sites of $S$ ($\partial S$) and
\[
{\bar P}_S=\inf_{\{-a\le U_\br'\le a\},-v\le w\le v}\prod_{\br\in S}P(U_\br'+w)
\ .
\]
${\bar P}_S|\partial S|$ is the contribution of the boundary of a cube
and the integral is the integrated DOS on a cube $S$. The boundary term is substracted
because we have removed the boundary. In other words, the left-hand side of (\ref{bound}) is
the average DOS on the entire lattice, the right-hand side is the average DOS on the isolated 
cube $S$.

The value of the lower bound requires an adjustment of the still undetermined parameters 
$a$ and $v$. The integrated DOS $\int_{-v}^v\sum_{\br \in S}\rho_{S,\br}(U'+E)dE$  on $S$
is the number of eigenvalues on the interval $[-v,v]$ of the $S$--projected Hamiltonian 
$H_0-U'$. The projected Hamiltonian is an $|S|\times |S|$ Hermitean matrix with finite elements, 
whose eigenvalues are also finite. Thus, for a fixed $a$ we can choose a sufficiently large $v$
such that all eigenvalues of the projected Hamiltonian are inside the interval $[-v,v]$.
In this case the integrated DOS is $|S|$ and we get from (\ref{bound}) the inequality
\beq
\sum_{\br\in S}{\bar\rho}_\br\ge {\bar P}_S[|S|-|\partial S|] 
\ .
\label{bound2}
\eeq 
$S$ can always be chosen such that the size of the cube $|S|$ is larger than the size
$|\partial S|$ of its boundary. Then the right-hand side of (\ref{bound}) is strictly positive.
$v$ should not be too large, though, in order to avoid that ${\bar P}_S$ becomes
too small, assuming that a typical $P(U_\br)$ decays for large values.
The actual value of $P_S$ depends on the distribution and can be exponentially small.

The average DOS of the entire lattice is estimated by the sum over all cubes, normalized by 
its number $N$. Since all cubes have the same lower bound, this sum is bounded by the right-hand 
side of (\ref{bound2}). This indicates that our estimation works only for a macroscopic number of 
impurities, the case of a single impurity (\ref{2x2}) would always give a lower bound zero.

\section{Conclusion}

There is a crucial difference in terms of the average DOS: For a discrete distribution
the average DOS is non-zero only if the disorder potential is ``resonant'' with the
pure Green's function $G_0$, according to the second term in Eq. (\ref{LSE}). In particular,
a single impurity fails to create spectral weight at the Weyl node.
On the other hand, for a dense distribution of impurities, represented by a continuous random potential, 
there is always a non-vanishing average DOS due to inter-impurity bound states, provided that the values 
of $U_\br$ cover the entire spectrum of $H_0$. 
  
The existence of a critical disorder strength $\gamma_c$, as indicated by the self-consistent
approximation in Eq. (\ref{scba2}), contradicts the existence of a lower non-zero bound of 
the average DOS in Sect. \ref{sect:lower_bound}. Therefore, the self-consistent calculation
is not sufficiently accurate to describe the transport properties of the 3D Weyl semimetal
properly. Since the lower bound of the average DOS is only a qualitative, although rigorous, estimation, 
still a reliable approximation is necessary to obtain an approximative value for the
average DOS. The exact result obtained for the Cauchy-Lorentz distribution in Sect.
\ref{sect:CLD} gives only a hint, because this distribution is not generic. A possible
option is a $N^{-\alpha}$ expansion with non-integer $\alpha$ \cite{ziegler83}.  
 

\vskip0.3cm

Acknowledgment: This work was supported by a grant of the Julian Schwinger Foundation.

\end{document}


\title{{\it Supplementary Material}:\\
Short note on the density of states in 3D Weyl semimetals
}

\author{K. Ziegler and A. Sinner}
\affiliation{
Institut f\"ur Physik, Universit\"at Augsburg,
D-86135 Augsburg, Germany}
\date{\today}

\maketitle

\section{Lower bound of the average density of states}
\label{sect:lower_bound}

The main idea of calculating the DOS of an infinite system is to divide the large system into 
smaller (finite) cubes, calculate the DOS of these smaller
cubes and estimate the contribution of the boundaries between them. This concept was 
developed for a periodic lattice by Ledermann \cite{ledermann44}. Later it was extended 
to estimate the average DOS of a tight-binding model with random potential by Wegner \cite{wegner81}, 
using the relation between
the DOS and the integrated DOS. The calculational advantage of using a finite cube is its discrete
spectrum. Then the corresponding DOS is a sum of Dirac Delta functions (or poles of the corresponding
Green's function), which can be studied, for instance, by averaging with respect to a continuous 
disorder distribution. Then the Dirac Delta functions contribute to the average with their 
spectral weights.

This concept can be generalized by introducing a generating function for the local DOS,  
which is the phase of a unimodular function \cite{ziegler87,ziegler88,ziegler98}. The phase has special 
properties under the change of the matrix elements of the underlying tight-binding Hamiltonian, 
which leads to a flexible method for estimating the average DOS. 

For a diagonal matrix $U$ and a short-range tight-binding matrix $H_0$ with lattice sites $\{\br\}$
there is a generating function 
\beq
F_\Lambda=i\log\left[\frac{\det(H_0-U-i\epsilon)}{\det(H_0-U+i\epsilon)}\right]
\ \ \ {\rm with}\ 
\rho_\br=\frac{1}{2\pi}\frac{\partial F_\Lambda}{\partial U_\br}
\ \ \ (\epsilon>0)
\label{def_F}
\eeq
for the local density of states $\rho_\br$ on the lattice $\Lambda$. 
The specific form of $H_0$ is not important for the following discussion, as long as it
is short ranged. The latter is crucial because it allows us to obtain a sufficiently
thin surface to disconnected cubes on the lattice $\Lambda$. Whether $H_0$ is a symmetric
tight-binding Hamiltonian or a discrete Dirac operator with spinor states does not affect
the validity of the approach.

$F_\Lambda$ has some remarkable properties: It is real, since the argument of the logarithm 
is unimodular, and it is an increasing function for any $U_\br$, since the DOS is non-negative.
$F_\Lambda$ is bounded for the shift of a single variable $U_\br'\to U_\br$ as
\beq
0\le F_\Lambda(U_\br)-F_\Lambda(U_\br')\le 2\pi\ \ (U_\br > U_\br')
\label{diff1}
\eeq
and for $n$ shifted variables 
$U_{\br_1}',U_{\br_2}',...,U_{\br_n}'\to U_{\br_1},U_{\br_2},...,U_{\br_n}$ it is bounded as
\beq
0\le F_\Lambda(U_{\br_1},U_{\br_2},...,U_{\br_n})-F_\Lambda(U_{\br_1}',U_{\br_2}',...,U_{\br_n}')
\le 2\pi n\ \ (U_{\br_j} > U_{\br_j}')
\ .
\label{diff2}
\eeq
$F_\Lambda$ is additive on the lattice in the limit $U_\br\to\infty$ on $\partial S$: 
\beq
\lim_{U_\br\to\infty,\ \br\in \partial S}F_\Lambda= F_S+F_{S''}
\label{add1}
\eeq
for a cube $S$ with the boundary $\partial S$ and the complement $S''$ outside $S\cup\partial S$.
$F_S$ ($F_{S''}$) is the function $F_\Lambda$ with $H_0-U\pm i\epsilon$ projected onto the subspace 
$S$ ($S''$).
The combination of (\ref{diff2}) and (\ref{add1}) implies
\beq
F_S+F_{S''}-2\pi |\partial S|\le F_\Lambda\le F_S+F_{S''} 
\ ,
\label{ineq1}
\eeq
where $|\partial S|$ is the number of lattice sites in $\partial S$.

Before we discuss the lower bound of the average DOS, an interpretation of the generating
function $F_\Lambda$ might be useful. According to its definition in Eq. (\ref{def_F})
$F_\Lambda/2$ is the phase of the determinant of $H_0-U-i\epsilon$. If we increase a single
$U_\br$ this phase also increases, as indicated by (\ref{diff1}). Since this happens for
the increase of any of elements of $U$, the phase changes add up and the shift
$U_{\br_1}',U_{\br_2}',...,U_{\br_n}'\to U_{\br_1},U_{\br_2},...,U_{\br_n}$ provides a winding
number of the determinant. The change of all elements of $U$ by a constant $E$ leads to
the integrated DOS on the interval $[0,E]$:
\beq
N(0,E)=\int_0^E\sum_\br\rho_\br(U+y)dy
\ ,
\eeq
which is the number of eigenvalues of $H_0-U$ on the interval $[0,E]$. In other words, the
winding number of a global change of $U$ is equal to the number of eigenstates on the interval
of the change.  

Now we return to the average DOS
\beq
\sum_{\br\in S}{\bar \rho}_\br
:=\sum_{\br\in S}\int\rho_\br(U) \prod_{\br\in \Lambda}P(U_\br)dU_\br
=\frac{1}{2\pi}\int\sum_{\br\in S}\frac{\partial F_\Lambda}{\partial U_\br} 
\prod_{\br\in \Lambda}P(U_\br)dU_\br
\ .
\label{ados}
\eeq
The distribution density on $S$ can be written as an integral transform
\beq
\prod_{\br\in S}P(U_\br)
=P'(U\Pi_S)\int_{-v}^v \prod_{\br\in S}P(U_\br-u)du
\ ,
\label{int-tr}
\eeq
where $\Pi_S$ is the projector onto $S$.
This gives us for the right-hand side of (\ref{ados})
\[
\frac{1}{2\pi}\int\sum_{\br\in S}\frac{\partial F_\Lambda(U)}{\partial U_\br} 
\int_{-v}^v\prod_{\br\in S}P(U_\br-u)duP'(U\Pi_S)\prod_{\br\in S}dU_\br \prod_{\br\notin S}P(U_\br)dU_\br
\]
and with the new integration variable $U_\br'=U_\br-u$ we have
\[
=\frac{1}{2\pi}\int\int_{-v}^v\sum_{\br\in S}\frac{\partial F_\Lambda(U'+u\Pi_S)}{\partial U_\br}
P'((U'+u)\Pi_S)du 
\prod_{\br\in S}P(U_\br')dU_\br'\prod_{\br\notin S}P(U_\br)dU_\br
\]
\beq
\ge \frac{1}{2\pi}\int\inf_{-v\le w\le v}P'((U'+w)\Pi_S)
\int_{-v}^v\sum_{\br\in S}\frac{\partial F_\Lambda(U'+u\Pi_S)}{\partial U_\br}du 
\prod_{\br\in S}P(U_\br')dU_\br'\prod_{\br\notin S}P(U_\br)dU_\br
\eeq
With the relation
\[
\int_{-v}^v\sum_{\br\in S}\frac{\partial F_\Lambda(U'+u\Pi_S)}{\partial U_\br}du
=\int_{-v}^v\frac{\partial F_\Lambda(U'+u\Pi_S)}{\partial u}du
=F_\Lambda(U'+v\Pi_S)-F_\Lambda(U'-v\Pi_S)
\]
we obtain
\beq
\sum_{\br\in S}{\bar \rho}_\br
\ge\frac{1}{2\pi}\int [F_\Lambda(U'+v\Pi_S)-F_\Lambda(U'-v\Pi_S)]
\inf_{-v\le w\le v}P'((U'+w)\Pi_S)\prod_{\br\in S}P(U_\br')dU_\br'\prod_{\br\notin S}P(U_\br)dU_\br
\ .
\eeq
Now we apply the inequalities (\ref{ineq1}) and get a lower bound
\beq
\sum_{\br\in S}{\bar \rho}_\br
\ge\frac{1}{2\pi}\int [F_S(U'+v)-F_S(U'-v)-2\pi|\partial S|]\inf_{-v\le w\le v}P'((U'+w)\Pi_S)
\prod_{\br\in S}P(U_\br')dU_\br'
\ ,
\eeq
where the integration outside of $S$ has been performed, since the integrand
does not depend on $U_\br$ for $\br\notin S$. Next, we estimate the integral as
\[
\frac{1}{2\pi}\int [F_S(U'+v)-F_S(U'-v)-2\pi|\partial S|]\inf_{-v\le w\le v}P'((U'+w)\Pi_S)
\prod_{\br\in S}P(U_\br')dU_\br'
\]
\beq
\ge 
\inf_{-a\le U_\br'\le a,\ \br\in S}[F_S(U'+v)-F_S(U'-v)-2\pi|\partial S|]
\frac{1}{2\pi}\inf_{-v\le w\le v}P'((U'+w)\Pi_S) 
\ .
\eeq
$F_S(U'+v)-F_S(U'-v)$ is the integrated DOS on the cube $S$
\[
\int_{-v}^v\sum_{\br \in S}\rho_{S,\br}(U'+E)dE
\ ,
\]
which counts the number of eigenstates of the $|S|\times|S|$--matrix
$\Pi_S(H_0-U')\Pi_S$ on the interval $[-v,v]$. Finally, from Eq. (\ref{int-tr}) we get
\beq
P'((U'+w)\Pi_S)=\frac{\prod_{\br\in S}P(U_\br'+w)}{\int_{-v}^v\prod_{\br\in S}P(U_\br'+w-u)du}
\ge \prod_{\br\in S}P(U_\br'+w)
\ ,
\label{fin_in}
\eeq
which gives inequality (19) of the Letter.

\section{Derivation of properties (\ref{diff2}) and (\ref{add1})}
\label{sect:properties}

As discussed above, we obtain a lower bound of the average DOS essentially through properties  
(\ref{diff2}) and (\ref{add1}) of the generating function $F_\Lambda$.
These properties were discussed previously in Refs. \cite{ziegler87}--\cite{ziegler98}
but for a consistent notation we summarize them in the following.

\subsection{Property (\ref{diff2})}

The inequality (\ref{diff1}) for one shifted variable is directly related to the Lippmann-Schwinger 
equation for a single impurity via
\beq
F_\Lambda(U_\br')-F_\Lambda(U_\br'') = 2\pi\int_{U_\br''}^{U_\br'} \rho_\br(U_\br)dU_\br
=-i\int_{U_\br''}^{U_\br'}
\left[\frac{1}{1/\gamma_\br^*-U_\br}-\frac{1}{1/\gamma_\br-U_\br}\right]dU_\br\le 2\pi
\ ,
\label{diff1a}
\eeq
where $\gamma_\br=G_{0,\br\br}$ is the spatial diagonal element of the Green's function. The integral 
is also non-negative because the imaginary part of $\gamma_\br$ is positive for $\epsilon>0$.
A special case is that of Weyl particles because of $\gamma_\br\sim 0$. Then the above expression is
always zero for finite $U_\br$, $U_\br'$ because the pole of the integrand is at infinity, as mention in
the Letter.

Then we apply two times (\ref{diff1a}) to obtain for two shifted variables 
\[
0\le F_\Lambda(U_{\br_1},U_{\br_2})-F_\Lambda(U_{\br_1}',U_{\br_2}')
= F_\Lambda(U_{\br_1},U_{\br_2})-F_\Lambda(U_{\br_1}',U_{\br_2})
+F_\Lambda(U_{\br_1}',U_{\br_2})-F_\Lambda(U_{\br_1}',U_{\br_2}')
\le 4\pi
\ .
\]
This procedure can be repeated $n$ times for $n$ shifted variables to give (\ref{diff2}). 
The result is justified by complete induction: Suppose (\ref{diff2}) is correct. Then we get for $n+1$
\[
F_\Lambda(U_{\br_1},U_{\br_2},...,U_{\br_{n+1}})-F_\Lambda(U_{\br_1}',U_{\br_2}',...,U_{\br_{n+1}}')
=F_\Lambda(U_{\br_1},U_{\br_2},...,U_{\br_{n+1}})-F_\Lambda(U_{\br_1}',U_{\br_2}',...,U_{\br_n}',U_{\br_{n+1}})
\]
\[
+F_\Lambda(U_{\br_1}',U_{\br_2}',...,U_{\br_n}',U_{\br_{n+1}})
-F_\Lambda(U_{\br_1}',U_{\br_2}',...,U_{\br_n}',U_{\br_{n+1}}')
\le 2\pi n +2\pi=2\pi(n+1)
\ .
\]

\subsection{Property (\ref{add1})}

The relation (\ref{add1}) can be derived by splitting $U=U_{S\cup S''}+U_{\partial S}$ with
the projectors $\Pi_S$, $\Pi_{S''}$ and $\Pi_{\partial S}$ onto $S$, $S''$ and $\partial S$,
respectively:
\[
U_{S\cup S''}=\Pi_SU\Pi_S+\Pi_{S''}U\Pi_{S''}-\Pi_S-\Pi_{S''}
\ ,\ \ 
U_{\partial S}=\Pi_{\partial S}U\Pi_{\partial S}+\Pi_S+\Pi_{S''}
\ .
\]
Then we obtain the following equations
\[
\frac{\det(H_0-U-i\epsilon)}{\det(H_0-U+i\epsilon)}
=\frac{\det(H_0-U_{S\cup S''}-U_{\partial S}-i\epsilon)}
{\det(H_0-U_{S\cup S''}-U_{\partial S}+i\epsilon)}
\]
\beq
=\frac{\det\{U_{\partial S}^{1/2}[
U_{\partial S}^{-1/2}(H_0-U_{S\cup S''}-i\epsilon)U_{\partial S}^{-1/2}-{\bf 1}]U_{\partial S}^{1/2}\}}
{\det\{U_{\partial S}^{1/2}[
U_{\partial S}^{-1/2}(H_0-U_{S\cup S''}+i\epsilon)U_{\partial S}^{-1/2}-{\bf 1}]U_{\partial S}^{1/2}\}}
=\frac{\det[
U_{\partial S}^{-1/2}(H_0-U_{S\cup S''}-i\epsilon)U_{\partial S}^{-1/2}-{\bf 1}]}
{\det[
U_{\partial S}^{-1/2}(H_0-U_{S\cup S''}+i\epsilon)U_{\partial S}^{-1/2}-{\bf 1}]}
\ .
\eeq
The limit $U_\br\to\infty$ on $\partial S$ gives
\beq
\lim_{U_\br\to\infty,\ \br\in \partial S}U_{\partial S}^{-1/2}=\Pi_S+\Pi_{S''}
\ .
\eeq
Since $\partial S$ separates two regions $S$ and $S''$ on the lattice
with $\Pi_SH_0\Pi_{S''}=0$, this leads to
\[
\lim_{U_\br\to\infty,\ \br\in \partial S} U_{\partial S}^{-1/2}(H_0-U_{S\cup S''}\pm i\epsilon)U_{\partial S}^{-1/2}
=(\Pi_S+\Pi_{S''})(H_0-U_{S\cup S''}\pm i\epsilon)(\Pi_S+\Pi_{S''})
\] 
and, because of $\Pi_SH_0\Pi_{S''}=0$, we get
\beq
=\Pi_{S}(H_0-U\pm i\epsilon)\Pi_{S}+\Pi_{S''}(H_0-U\pm i\epsilon)\Pi_{S''}
\ .
\eeq
For the ratio of determinants we have
\beq
\lim_{U_\br\to\infty,\ \br\in \partial S}
\frac{\det(H_0-U-i\epsilon)}{\det(H_0-U+i\epsilon)}
=\frac{\det_S(H_0-U-i\epsilon)}{\det_S(H_0-U+i\epsilon)}
\frac{\det_{S''}(H_0-U-i\epsilon)}{\det_{S''}(H_0-U+i\epsilon)}
\ ,
\eeq
where the index of the determinants refers to the projection of the matrix.
Inserting this result into $F_\Lambda$ gives Eq. (\ref{add1}).